\documentstyle[prl,aps]{revtex}

\begin{document}
\author{Jian Qi Shen $^{1,}$$^{2}$ \footnote{E-mail address: jqshen@coer.zju.edu.cn}}
\address{$^{1}$   Centre for Optical
and Electromagnetic Research, Joint Research Centre of Photonics
of The Royal Institute of Technology (Sweden) and Zhejiang
University, Zhejiang University, Hangzhou Yuquan 310027, P.R. China\\
$^{2}$ Zhejiang Institute of Modern Physics and Department of
Physics, Zhejiang University, Hangzhou 310027, P.R. China}
\date{\today}
\title{Automatical generation of nonadiabatic conditional geometric phase shift \\via
a noncoplanar fiber system} \maketitle

\begin{abstract}
A new scheme of realizing the {\it nonadiabatic} conditional
geometric phase shift via a noncoplanar (and coiled) fiber system
is presented in this Letter. It is shown that the effective
Hamiltonian that describes the interaction of polarized photons
with the fiber medium is just the Wang-Keiji type of Hamiltonian.
This, therefore, means that the coiled fiber system may be an
ideal implementation of realizing the nonadiabatic geometric phase
gates for the topological quantum computation. The remarkable
feature of the present method is that it can {\it automatically}
meet the conditions and requirements proposed in the Wang-Keiji
scheme: (i) in the coiled fiber system, the dynamical phase of
photon wavefunction caused by the interaction Hamiltonian
automatically vanishes; (ii) the Wang-Keiji requirement for the
parameters in the Wang-Keiji Hamiltonian can be exactly satisfied
automatically in the fiber system; (iii) the conditional initial
state can be easily achieved by manipulating the initial wave
vector of polarized photons. Due to these three advantages, the
coiled fiber system may be a potentially practical way of
achieving the nonadiabatic conditional geometric phase shift (and
hence the nonadiabatic geometric quantum gates).

 PACS number(s): 03.65.Vf, 03.67.Lx

\end{abstract}

Recently, much attention has been attracted to the area of the
topological quantum computation both theoretically and
experimentally\cite{Zanardi,Margolin,Jones,Wang2,Wang1}. Several
basic ideas for realizing the adiabatic geometric quantum
computation by using nuclear magnetic resonance (NMR)\cite{Jones},
superconducting nanocircuits\cite{Falci} and trapped
ions\cite{Duan} were suggested. An experimental realization of the
conditional adiabatic phase shift with NMR technique was performed
by Jones {\it et al.}\cite{Jones}. It is well known that due to
its topological and global nature\cite{Berry}, the geometric phase
of a spinning particle will not be affected by the random
fluctuation arising in the evolution path\cite{Zanardi,Jones}.
This, therefore, means that the geometric phase shift gates
(should such exist) may be robust with respect to certain types of
operational errors. Since in the quantum computer a fault-tolerant
quantum logic gate is particularly essential for realizing a
quantum information processor, the quantum computation based on
this kind of geometric phase shift gates will inevitably become an
ideal scenario to this purpose\cite{Wang2}. Indeed, in the
experiments performed at the end of last century, such conditional
geometric phase shift gates was realized via NMR under adiabatic
conditions\cite{Cory,Gershenfeld,Jones2,Jones3}. Here the
adiabatic requirement results from the fact that only when the
time-dependent parameters of the Hamiltonian of qubits vary
extremely slowly will the result of quantum computation be
exact\cite{Jones}. Once if the adiabatic requirement is satisfied,
we meet, however, with the problem of quantum decoherence caused
by the inevitable interaction between the qubits (spin) and the
environments (such as a noise field, thermal reservoir and bath).
One reason that quantum computers will be difficult to build is
{\it decoherence}. In the process of decoherence, some qubit or
qubits of the computation become entangled with the environment,
thus in effect collapsing the state of the quantum
computer\cite{Shor}. The existence of such decoherence effects
demands that every operation in quantum computation process (in
which the maintenance of coherence over a large number of states
is essential) should be completed within the decoherence time. If
not, the decoherence error will unavoidably increase. To avoid the
decoherence error, one can choose to change the parameters of
Hamiltonian rapidly. But, as stated above, such a choice may break
the adiabatic requirement and therefore leads to the
nonadiabaticity error in the results of quantum
computation\cite{Jones}. Hence, it seems that there is inescapably
a conflict between the requirement of adiabatic condition (to
avoid the severe distortion from the nonadiabaticity to the
results) and the removal of decoherence effects. Because of this
conflict, such conditional geometric phase shift gates (both
adiabatic and nonadiabatic) has the limitation which restricts
greatly the scope of its applications. In a strict sense, this
scenario may be not quite beneficial to the practical quantum
computation.

In order to overcome this conflict, more recently, Wang and Keiji
suggested an easy scheme to exactly control the state evolution on
the cone and to make the geometric phase shift nonadiabatically
with NMR, which will not distort a quibt with its initial state in
the evolution and can therefore possibly get the exact results of
quantum computation\cite{Wang2}. It is well known that a spinning
charged particle in a time-dependent magnetic field ({\it e.g.}, a
field rotating around \^{z}-axis with a certain angular speed)
will acquire a geometric phase. The Hamiltonian of such a
time-dependent quantum system is of the form
$H_{0}=\omega_{1}(\cos\gamma t S_{x}+\sin\gamma t
S_{y})+\omega_{0}S_{z}$, where $\omega_{1}$ denotes the amplitude
of the horizontal magnetic field that is rotating around
\^{z}-axis with a constant angular speed $\gamma$, and
$\omega_{0}$ the amplitude of the vertical magnetic field. To
achieve the effective nonadiabatic conditional geometric phase
shift that will not lead to the nonadiabaticity error in the
results of quantum computation, Wang and Keiji introduced an
additional vertical static magnetic field, the amplitude of which
is equal to the rotating angular frequency $\gamma$ of the
horizontal field $\omega_{1}$\cite{Wang2}. They showed that such a
method can make the spinning particle evolve nonadiabatically in
the conical evolution process. This means that the conditional
geometric phase shift gates can be run nonadiabatically and hence
the topological quantum computation, which will not be affected by
both decoherence error and nonadiabaticity error, can possibly be
realized. After introducing the additional vertical magnetic
field, the Hamiltonian of the NMR system that Wang and Keiji
considered takes the form\cite{Wang2}
\begin{equation}
H(t)=\omega_{1}(\cos\gamma t S_{x}+\sin\gamma t
S_{y})+(\omega_{0}+\gamma)S_{z}. \label{Wang}
\end{equation}
In the present Letter, we refer to it as the Wang-Keiji
Hamiltonian. In the Wang-Keiji scheme\cite{Wang2}, the conditional
initial state
$\psi(0)=\cos(\theta/2)|\uparrow\rangle+\sin(\theta/2)|\downarrow\rangle$,
where $\theta$ is defined by $\theta=\arcsin
(\omega_{1}/\sqrt{\omega_{0}^{2}+\omega_{1}^{2}})$. This initial
state was created by a sequence of operations (the so-called S
operation) in the Wang-Keiji scheme\cite{Wang2}. However, in our
scenario, it may be achieved by a more convenient (or immediate)
way ({\it i.e.}, only controlling the wave vector of polarized
photons by the coiled fiber), which will be discussed further in
this Letter. In what follows, we will consider the possibility of
obtaining the nonadiabatic conditional geometric phase shift of
the polarized photons inside the curved fiber. It will be shown
that the coiled fiber system can automatically meet all the
conditions required in the Wang-Keiji NMR system, and it may
therefore be an ideal scenario to achieve the nonadiabatic
conditional geometric phase shift (and even hence the nonadiabatic
phase gates).

Assume that the noncoplanarly curved fiber can be viewed as a
perfect guide, namely, (i) the fiber does not alter the magnitude
but the direction of the photon wave vector ${\bf k}$ in the
evolution process, (ii) the wave vector ${\bf k}$ of photons can
be said to be always along the tangent to the curved guide (path)
at each point at arbitrary time in the evolution process inside
the fiber, and (iii) the photon helicity, $h={\bf k}\cdot{\bf
S}/k$, is conserved inside the curved fiber. It should be noted
that these assumptions have also been implied in the early work of
the adiabatic evolution of photons in the fiber by Chiao {\it et
al.}\cite{Chiao,Tomita}. According to the Liouville-Von Neumann
equation ${\partial h(t)}/{\partial t}-i[h(t),H(t)]=0$ (in the
unit $\hbar=1$)\cite{Lewis}, keeping in mind the first and third
assumptions, one can obtain the effective Hamiltonian $H$ that
describes the interaction between the photons and the curved
fiber, and consequently writes the time-dependent Schr\"{o}dinger
equation as follows\cite{Shen}
\begin{equation}
i\frac{\partial \left| \sigma ,{\bf{k}}(t)\right\rangle }{\partial
t}=\frac{{\bf{k}}(t)\times \dot{\bf{k}}(t)}{k^{2}}\cdot
{\bf{S}}\left|\sigma ,{\bf{k}}(t)\right\rangle,     \label{eqA3}
\end{equation}
where dot denotes the derivative with respect to time. $\sigma$ is
the eigenvalue of photon helicity, and $k$ the magnitude of ${\bf
k}$, {\it i.e.}, $k=|{\bf k}|$. Eq.(\ref{eqA3}) governs the wave
propagation of circularly polarized photons in the noncoplanar
fiber. Note that seen from Eq.(\ref{eqA3}), the propagation of the
photon inside a curved fiber can be described by a spin model, the
expression for the ``magnetic field'' of which is
${{\bf{k}}(t)\times \dot{\bf{k}}(t)}/{k^{2}}$. Although by using
the Lewis-Riesenfeld invariant theory\cite{Lewis}, we can exactly
solve the solutions of Eq.(\ref{eqA3}), where the fiber is
arbitrarily noncoplanarly curved, here we consider only the case
of the exact conical evolution ({\it i.e.}, the polar angle
$\theta$ is constant and the azimuthal angle $\varphi=\gamma t$).
Such a choice will lead to a convenient comparison of our results
with those obtained by Wang and Keiji\cite{Wang2}. In accordance
with the second assumption made above, the wave vector of photons
can characterize the geometric shape of the curved fiber. In order
to restrict the polar angle $\theta$ to a constant number, we
should consider a coiled fiber, where the wave vector of polarized
photons moving along a helical fiber can be written in the form $
{\bf k}=k(\sin\theta\cos\gamma t, \sin\theta\sin\gamma t,
\cos\theta)$. Note that here $\gamma$ is the rotating frequency of
photon moving on the helicoid, which is determined by the
geometric shape, {\it i.e.}, $\gamma=2\pi c/[n\sqrt{d^{2}+(4\pi
a)^{2}}]$, where $d$ and $a$ denote the pitch length and the
radius of the helix, respectively, and $n$ is the optical
refractive index of the fiber. Thus the phenomenological
``magnetic field'' is ${{\bf{k}}(t)\times
\dot{\bf{k}}(t)}/{k^{2}}=\gamma\sin\theta (-\cos\theta \cos\gamma
t, -\cos\theta\sin \gamma t, \sin\theta)$, and the Hamiltonian of
Schr\"{o}dinger equation (\ref{eqA3}) is rewritten
\begin{equation}
H(t)=-\gamma\sin\theta\cos\theta\left(\cos\gamma t
S_{x}+\sin\gamma t S_{y}\right)+\gamma\sin^{2}\theta S_{z}.
\label{Wang2}
\end{equation}
Comparison of (\ref{Wang2}) with (\ref{Wang}) shows that the
Hamiltonian (\ref{Wang2}) is just of a Wang-Keiji type. Moreover,
it is a special case of Wang-Keiji Hamiltonian, the parameters
$\theta$ and $\gamma$ of which satisfy the following relation
\begin{equation}
\omega_{1}^{2}+\omega_{0}^{2}+\gamma \omega_{0}=0,
\label{relation}
\end{equation}
where $\omega_{1}$ and $\omega_{0}$ are so defined that the polar
angle $\theta=\arctan(\omega_{1}/\omega_{0})$. The exact solution
of the time-dependent Schr\"{o}dinger equation Eq.(\ref{eqA3}) is
given
\begin{equation}
|\sigma, {\bf
k}(t)\rangle=\exp\left[\frac{1}{i}\phi_{\sigma}(t)\right]V(t)|\sigma\rangle,
\label{solution}
\end{equation}
where $|\sigma\rangle$ is the eigenstate of the third component of
photon spin, {\it i.e.},
$S_{z}|\sigma\rangle=\sigma|\sigma\rangle$, and the time-dependent
operator $ V(t)=\exp\left[\beta(t)
S_{+}-\beta^{\ast}(t)S_{-}\right]$\cite{Shen2}. Here
$\beta(t)=-(\theta/2)\exp(-i\gamma t)$,
$\beta^{\ast}(t)=-(\theta/2)\exp(i\gamma t)$, and
$S_{\pm}=S_{x}\pm iS_{y}$. The time-dependent phase of the
solution (\ref{solution}) is
$\phi_{\sigma}(t)=\int^{t}_{0}\langle\sigma|V^{\dagger}(t')H(t')V(t')-V^{\dagger}(t')i\partial
V(t')/\partial t'|\sigma\rangle {\rm d}t'$. Further calculation
shows that the noncyclic geometric phase is $\phi^{(\rm
g)}_{\sigma}(t)\equiv
-\int^{t}_{0}\langle\sigma|V^{\dagger}(t')i\partial V(t')/\partial
t'|\sigma\rangle {\rm d}t'=\sigma\gamma\left(1-\cos\theta
\right)t$. It should be noted that the dynamical phase resulting
from the effective Hamiltonian (\ref{Wang2}) is vanishing, {\it
i.e.},
\begin{equation}
\phi^{(\rm d)}_{\sigma}(t)\equiv
\int^{t}_{0}\langle\sigma|V^{\dagger}(t')H(t')V(t')|\sigma\rangle
{\rm d}t'=0.   \label{dynamical}
\end{equation}
So, the only retained phase in the time-dependent total phase,
$\phi_{\sigma}(t)$, of the solution (\ref{solution}) is the
noncyclic geometric phase $\phi^{(\rm g)}_{\sigma}(t)$. The reason
for the absence of the dynamical phase associated with the
Hamiltonian (\ref{Wang2}) is as follows: the ``magnetic field''
${{\bf{k}}(t)\times \dot{\bf{k}}(t)}/{k^{2}}$ gives rise to a
Lorentz type of force, which does not alter the magnitude of the
wave vector of photons in the curved fiber: specifically, it
follows from the Liouville-Von Neumann equation that the motion of
equation of photons inside the fiber is of the form $
\dot{\bf{k}}+{\bf{k}}\times \left({{\bf{k}}\times
\dot{\bf{k}}}/{k^{2}}\right)=0$, which can be readily shown to be
an identity. It is apparently seen that ${\bf{k}}\times
\left({{\bf{k}}\times \dot{\bf{k}}}/{k^{2}}\right)$ is just an
expression for a Lorentz type of force, and the ``magnetic field''
${{\bf{k}}(t)\times \dot{\bf{k}}(t)}/{k^{2}}$ is always
perpendicular to the photon wave vector. This, therefore, implies
that it has no contribution to the dynamical phase other than to
the geometric phase of the photon wavefunction (\ref{solution}).
This property ({\it i.e.}, $\phi^{(\rm d)}_{\sigma}(t)=0$) is very
fortunate for the topological quantum computation\cite{Wang2}.

 Now let us discuss the initial state of the solution (\ref{solution}). If $t=0$,
 the initial values of the parameters of $V$ are
 $\beta(0)=\beta^{\ast}(0)=-{\theta}/{2}$, and in consequence
$ V(0)=\exp\left(-i\theta S_{y}\right)$. If we assume that the
eigenvalue of the helicity of the applied circularly polarized
photons is $+1$, then it follows from (\ref{solution}) that the
initial state is expressed as
\begin{equation}
\psi(0)=V(0)|\uparrow\rangle=\cos\left(\frac{\theta}{2}\right)|\uparrow\rangle
+\sin\left(\frac{\theta}{2}\right)|\downarrow\rangle,
\label{initial}
\end{equation}
where $|\uparrow\rangle$ and $|\downarrow\rangle$ represent the
eigenstates of $S_{z}$ corresponding to the eigenvalues
$\sigma=\pm 1$, respectively. In the meanwhile, we can calculate
the initial wave vector of photons, which is $k_{x}=k\sin\theta,
k_{y}=0, k_{z}=k\cos\theta$. If at $t=0$, we restrict the wave
vector of the polarized photonic state with the helicity
eigenvalue $\sigma=+1$ to the \^{x}-\^{z} plane and let it agree
with the relation $k_{x}/k_{z}=\tan \theta$, then at least in
principle the conditional initial state $\psi(0)$ of photons will
be automatically achieved. In a word, it follows that the coiled
fiber by itself can guide the polarized photon into $\psi(0)$.

The above coiled fiber system to realize the nonadiabatic
conditional geometric phase shift of polarized photons may offer
the following three advantages:

(i) Both the propagation of photons inside a coiled fiber and the
interaction of the photon spin with the effective ``magnetic
field'' can be described by the Wang-Keiji type of Hamiltonian.
That is, the fiber system by itself satisfies the conditions of
Wang-Keiji scheme and can therefore generate the nonadiabatic
conditional geometric phase shift.

(ii) Fortunately for the nonadiabatic geometric phase shift gates,
the dynamical phase in the fiber system resulting from such
Wang-Keiji Hamiltonian (\ref{Wang2}) vanishes. In the Wang-Keiji
NMR scheme\cite{Wang2}, however, in order to remove the dynamical
phase acquired by qubits they should let the state evolve on the
dynamical phase path by choosing certain specific value of the
rotating speed $\gamma$ of the horizontal field (and also the
additional field $\gamma$). For this task, they required
$\gamma=-(\omega_{1}^{2}+\omega_{0}^{2})/\omega_{0}$. Here, it is
very interesting to see that this Wang-Keiji requirement is just
Eq.(\ref{relation}) obtained above by us. This, therefore, means
that finally Wang and Keiji chose our special Wang-Keiji type of
Hamiltonian (\ref{Wang2}) in their work to avoid the dynamical
phase\cite{Wang2}. But in the NMR system, it may be somewhat
difficult to meet this Wang-Keiji requirement for $\gamma$ very
exactly, in particular for the case of the time-dependent
$\gamma$\cite{Wang2}. In the coiled fiber system, however, as
stated above, the Wang-Keiji requirement can be automatically
exactly satisfied, since the Hamiltonian (\ref{Wang2}) by itself
leads to the relation (\ref{relation}), {\it i.e.}, the Wang-Keiji
requirement for the rotating speed of the horizontal field and the
amplitude of the additional field can be naturally met.
Particularly, even for the case of the time-dependent $\gamma$,
the dynamical phase of photons resulting from the interaction
Hamiltonian (\ref{Wang2}) automatically vanishes, namely, the
formula (\ref{dynamical}) holds for the arbitrarily time-dependent
functions $\theta(t)$ and $\varphi(t)$. In other words, the
Wang-Keiji requirement is still exactly met for the case of the
time-dependent $\gamma$.

(iii) In the Wang-Keiji NMR system, the authors proposed a
nonadiabatic way ({\it i.e.}, a sequence of operations, which they
called as S operation) to create the conditional initial state
(\ref{initial}), which is required in realizing the nonadiabatic
conditional geometric phase shift gates\cite{Wang2}. However, in
the coiled fiber system, the conditional initial state
(\ref{initial}) can be easily prepared ({\it i.e.}, the polarized
photon can be guided into the conditional initial state by the
coiled system). In other words, only the requirement of the
appropriate initial wave vector of photons ({\it i.e.},
$k_{x}/k_{z}=\tan \theta$) of the polarized photons (say,
$\sigma=+1$) will automatically lead to the conditional initial
state (\ref{initial}).

Although the coiled fiber system has these advantages for
realizing the nonadiabatic conditional geometric phase shift, at
least for the present the nonadiabatic geometric phase gates may
not be easily realized experimentally by such a fiber scheme. At
present, a number of systems have been suggested as the potential
quantum computer models, including the NMR, trapped ions, cavity
quantum electrodynamics, quantum dots, superconducting quantum
interference and so on. But maybe for some certain technological
difficulties, in the literature, less attention is paid to the
polarized photon interference to achieve the (topological) quantum
phase gates (and hence the quantum computation). So, we think that
the coiled fiber scheme for the nonadiabatic conditional geometric
phase gates may not be more effective than the Wang-Keiji NMR
scheme in the practical applications. However, the present
discovery that the coiled fiber system can {\it automatically}
meet all the conditions required in the Wang-Keiji NMR scheme is
still impressive. For this reason, we hold that the scheme
presented here deserves further investigation.

To summarize, in order to make the conditional geometric phase
shift gates be run nonadiabatically, Wang and Keiji should
introduce a suitable additional vertical magnetic field in their
NMR system; in order to produce a conditional initial state, they
should use a sequence of operations (S operation) on the Bloch
sphere; in order to remove the dynamical phase arising from the
interaction Hamiltonian between the spinning magnetic moment and
the external magnetic fields, Wang and Keiji should require the
amplitudes of vertical and horizontal fields, the rotating
frequency of horizontal field and the strength of the introduced
additional field to exactly agree with a certain relation ({\it
i.e.}, the so-called Wang-Keiji requirement\cite{Wang2}).
Fortunately, all these conditions and requirements suggested in
the Wang-Keiji NMR system can be satisfied by itself in our scheme
of coiled fiber system. Thus, it may be believed that the scheme
of coiled fiber system might be an ideal scenario for realizing
the nonadiabatic conditional geometric gates, on which the
topological quantum computation (which may be robust with respect
to the decoherence and nonadiabaticity errors) is based. We hope
this coiled fiber scheme would be investigated experimentally in
the near future.

 \textbf{Acknowledgements}  This work was supported
partially by the National Natural Science Foundation of China
under Project No. $90101024$ and $60378037$. I thank X.B. Wang for
his drawing my attention to their related work on the topological
quantum computation.


\begin{references}
\bibitem{Zanardi} P. Zanardi and M. Rasetti, Phys. Lett. A {\bf
264}, 94 (1999).


\bibitem{Margolin} A.E. Margolin, V.I. Strazhev, and A.Y.
Tregubovich, arXiv: quant-ph/0102030 (unpublished); S.L. Zhu and
Z.D. Wang, Phys. Rev. A {\bf 66}, 042322 (2002); S.L. Zhu and Z.D.
Wang, Phys. Rev. Lett. {\bf 89}, 097902 (2002); V. Vedral, arXiv:
quant-ph/0212133 (unpublished).


\bibitem{Jones} J.A. Jones, V. Vedral, A. Ekert, and G.
Castagnoli, Nature {\bf 403}, 869 (2000).


\bibitem{Wang2} X.B. Wang and M. Keiji, Phys. Rev. Lett. {\bf 87},
097901 (2001); (E) Phys. Rev. Lett. {\bf 88}, 179901 (2002).

\bibitem{Wang1} X.B. Wang and M. Keiji, Phys. Rev. B {\bf 65},
172508 (2002).

\bibitem{Falci} G. Falci, R. Fazio, G.M. Palma {\it et al.},
Nature {\bf 407}, 355 (2000).

\bibitem{Duan} L.M. Duan, J.I. Cirac, and P. Zoller, Science {\bf
292}, 1695 (2001).

\bibitem{Berry} M.V. Berry, Proc. R. Soc. London, Ser. A {\bf 392},
45 (1984).

\bibitem{Cory} D.G. Cory, A.F. Fahmy, and T.F. Havel, Proc. Nat.
Acad. Sci. USA {\bf 94}, 1634 (1997).

\bibitem{Gershenfeld}  N.A. Gershenfeld and I.L. Chung, Science {\bf
275}, 350 (1997).

\bibitem{Jones2} J.A. Jones and M. Mosca, J. Chem. Phys. {\bf
109}, 1648 (1998).

\bibitem{Jones3} J.A. Jones, R.H. Hansen, and M. Mosca, J. Magn.
Reson. {\bf 135}, 353 (1998).

\bibitem{Shor} P.W. Shor, Phys. Rev. A {\bf 52}, R2493 (1995).

\bibitem{Chiao}  R.Y. Chiao and Y.S. Wu, Phys. Rev. Lett. {\bf
57}, 933 (1986).

\bibitem{Tomita}   A. Tomita and R.Y. Chiao, Phys. Rev. Lett. {\bf 57},
937 (1986).

\bibitem{Lewis}  H.R. Lewis and W.B. Riesenfeld, J. Math. Phys. {\bf
10}, 1458 (1969).

\bibitem{Shen} J.Q. Shen and L.H. Ma, Phys. Lett. A {\bf 308},
355 (2003).

\bibitem{Shen2} J.Q. Shen, H.Y. Zhu, and P. Chen, Eur. Phys. J. D
{\bf 23}, 305 (2003).

\end{references}
\end{document}